\documentclass[epj,nopacs]{svjour}

\usepackage{mathrsfs}
\usepackage{amssymb}
\usepackage{amsmath}
\usepackage{amssymb}
\usepackage{graphicx}
\usepackage{subfigure}

\begin{document}

\title{Spin glass approach to the feedback vertex set problem}

\author{
Hai-Jun Zhou \thanks{e-mail: \tt{zhouhj@itp.ac.cn}}
}

\institute{
State Key Laboratory of Theoretical Physics, Institute of Theoretical Physics, Chinese Academy of Sciences, Zhong-Guan-Cun East Road 55, Beijing 100190, China
}

\date{Received: date / Revised version: date}

\abstract{
  A feedback vertex set (FVS) of an undirected graph is a set of vertices
  that contains at least one vertex of each cycle of the graph.
  The feedback vertex set problem consists of constructing a FVS of
  size less than a certain given value.
  This combinatorial optimization problem has many practical
  applications,  but it is in the nondeterministic
  polynomial-complete class of worst-case computational complexity.
  In this paper we define a spin glass model
  for the FVS problem and then study this model
  on the ensemble of finite-connectivity random graphs.
  In our model the global cycle constraints
  are represented through the local constraints on all the
  edges of the graph, and they are then treated by distributed
  message-passing procedures such as
  belief propagation. Our belief propagation-guided decimation
  algorithm can construct nearly optimal
  feedback vertex sets for single random graph instances and
  regular lattices.
  We also design  a spin glass model for the FVS problem on a
  directed graph.
  Our work will be very useful for identifying the set of vertices that
  contribute most significantly to the dynamical complexity of a large
  networked system.
}

\maketitle

\section{Introduction}

The feedback vertex set (FVS) problem is a fundamental combinatorial
optimization problem
in the field of computation complexity. It is among the
first $21$ problems shown to be
nondeterministic polynomial-complete (NP-complete) by
Cook and Karp in the early 1970s
\cite{Cook-1971,Karp-1972,Garey-Johnson-1979}.
For an undirected graph, a FVS is a vertex set
which contains at least one vertex of every cycle of this graph.
In other words, after all the vertices in the FVS have been
removed, the remaining graph will be free of any
cycles (it is a forest, i.e., a collection of trees).
A FVS for a directed graph is similarly defined, namely such a set
should contain at least one vertex of every directed cycle of the
graph.
A feedback vertex set is also referred to
as a decycling set in some references \cite{Beineke-Vandell-1997}.

The FVS problem has wide practical applications, such as
deadlock recovery in operation systems and
combinatorial circuit design \cite{Zobel-1983},
dynamics of regulatory networks
\cite{Fiedler-etal-2013,Mochizuki-etal-2013}, and
network control and observation
\cite{Liu-Slotine-Barabasi-2011,Liu-Slotine-Barabasi-2013}.
For example, a dynamical system of two-body interactions can be
represented as a graph of vertices and edges.
Such a system can be divided into
a $`$boundary' (containing all the vertices of a FVS)
and an $`$interior' (containing all the other vertices).  Since
the interior contains no cycles, its dynamical behavior in principle is
completely determined by the states of the vertices in the boundary.
Therefore the dynamical behavior of the whole system can be
monitored through controlling the states of the vertices in the FVS.
For many practical purposes it is naturally very desirable
to construct a FVS that contains as few vertices as
possible.

Each vertex of the graph has a non-negative weight, and the
weight of a FVS is just the sum of the weights of its constituent
vertices. A FVS is referred to as an optimal (or minimum) one
if its weight is the global minimum value (denoted as
$W_0$) among all the possible feedback vertex sets of a given graph.
The goal of the FVS problem is to construct a
FVS of weight not exceeding a certain prescribed value, say $W^*$. The
difficulty of the FVS problem increases as the value
$W^*$ decreases. The most challenging issue is the minimum
FVS problem which corresponds to $W^* = W_0$.

Despite its theoretical and practical importance,
the FVS problem has not been much investigated by
the statistical physics community.
Cycles of all sizes need to be considered in the FVS problem
(see
\cite{Bianconi-Marsili-2005,Marinari-Monasson-Semerjian-2006,Marinari-Semerjian-2006,Marinari-Semerjian-VanKerrebroeck-2007,Bianconi-Gulbahce-2008} for some recent interesting papers on the cycle counting and 
construction problem).
One of the main obstacles is that
cycles are global structural properties of a graph.
One usually can not judge whether cycles are absent in a graph by
only looking at single vertices or edges.
This theoretical difficulty is solved in this work for
the FVS problem on undirected graphs.
We have found a simple way of representing the global
cycle constraints of the FVS problem through the local
constraints on all the edges of the graph. A spin glass model is
constructed for the FVS problem by defining an integer-valued state
variable on each vertex and then applying a local
constraint on each edge.
We study this spin glass model on the ensemble of finite-connectivity
random graphs by mean field theory, and then apply a
message-passing algorithm (inspired by this mean field theory) to single
random graph instances and  hyper-cubic regular lattices.
We find that our algorithm is able to construct nearly optimal
feedback vertex sets for single random graph instances and
regular lattice instances.

We also construct a similar spin glass model for the FVS problem on a
  directed graph. Detailed investigations on this second
  model will be carried out in a separate work.

This paper is organized as follows. In the next section we define the
FVS problem more precisely and introduce some graph concepts. In
section \ref{sec:model} the spin glass model for the FVS problem on
undirected graphs is introduced.
This spin glass model is analyzed by the replica-symmetric mean field
theory in section \ref{sec:RS} and by belief propagation-guided
decimation algorithm in section \ref{sec:BPD}. We conclude our
work in section \ref{sec:conclusion} and  discuss some possible
extensions.

\section{The undirected feedback vertex set problem}
\label{sec:FVS}

We consider an undirected and simple graph $G$ \cite{He-Liu-Wang-2009}.
There are $N$ vertices in the
graph, whose integer-valued indices
(generically denoted as $i, j, k, \ldots$)
range from $1$ to $N$.
There are $M$ edges in the graph, each of which connects two
different vertices. If there is an edge between two vertices $i$ and
$j$, this edge is then denoted as $(i, j)$.
The edges have no intrinsic directions,
therefore the graph is undirected.
There are no self-edges that connect a vertex  to itself, and
there is at most one edge between any pair of different vertices.

If there is an edge between a vertex $i$ and another
vertex $k$, then vertex $k$ is referred to as a neighbor of vertex $i$ and
$i$ a neighbor of $k$. The set of neighbors of a vertex $i$ is denoted as
$\partial i$ and the degree $d_i$ of vertex $i$ is just its number of attached
edges, namely $d_i \equiv |\partial i|$.

A path in a graph $G$ is a sequence of edges which connect a
sequence of vertices, for example a path
$$
(i, j_1), \ (j_1, j_2),\
\ldots, \ (j_{n-1}, j_{n}),\  (j_n, j)
$$
connecting vertex $i$ and $j$.
If the start and the end vertex of a path are the
same, such a path is referred to as a cycle.
A tree of graph $G$ is a connected subgraph that contains no cycles.

A feedback vertex set (FVS) of graph $G$ is a subset
$\Gamma$ of the $N$ vertices
such that if all the vertices of this set and the attached edges
are removed from $G$ the remaining graph will have no cycles and
simply be a collection of  tree components.
Therefore for each cycle of the graph $G$, at least one vertex on
this cycle is contained in the set $\Gamma$.

Constructing a FVS for a given graph is a rather easy task.
A simple recipe would be to repeatedly remove a randomly chosen
vertex from the graph until there is no cycle in the
graph.
However the optimization problem of constructing a FVS of the
global minimum weight
(a minimum feedback vertex set)
is extremely non-trivial. Indeed the
minimum FVS problem
is a combinatorial optimization problem in the
nondeterministic-polynomial-hard (NP-hard)
complexity
class \cite{Garey-Johnson-1979}. It is generally believed that
no deterministic sequential algorithm is able to
construct a minimum FVS for all
input graphs $G$ in a computing time that grows
only polynomially with
the number $N$ of vertices in $G$.

\section{Spin glass model}
\label{sec:model}

In this work we study the undirected FVS problem using statistical physics
methods. For a given large graph $G$,
the aim is to construct a subgraph that contains as many
vertices as possible but is free of cycles. Since cycles are not
necessarily local structures of a graph,  the requirement that the
subgraph should have no cycles is a very strong global
constraint on the property of the system.
An important first step of our statistical physics approach is
to turn the global cycle constraints into a set of local
constraints. This challenging task has been accomplished by the
following simple model construction.

First, let us define on each vertex $i$
a state variable $A_i$, which can take the value $A_i=0$, $A_i=i$ or
  $A_i = j \in \partial i$. Therefore the state $A_i$ of vertex $i$ can have
  $d_i + 2$ different choices and the state sets of different vertices
  are different.
If $A_i=0$ we say that vertex $i$ is un-occupied; if $A_i=i$ we say that vertex
$i$ is occupied and it is a root vertex (it has no
parent vertex);
if $A_i= j \in \partial i$ we say that
vertex $i$ is occupied and its parent vertex is $j$.
An edge $(i, j)$ of the graph $G$
is regarded as un-occupied if either $A_i=0$ or $A_j=0$,
otherwise it is regarded as  occupied. We realize that such
a vertex state variable $A_i$ has also been defined
in an earlier study of the
Steiner tree problem by Zecchina and co-workers \cite{Bayati-etal-2008,BaillyBechet-etal-2011,Biazzo-Braunstein-Zecchina-2012}
(in which $A_i$ is denoted as $p_i$ and each vertex $i$ has an additional
depth state variable $h_i$).

\begin{figure}
\begin{center}
\includegraphics[width=0.45\textwidth]{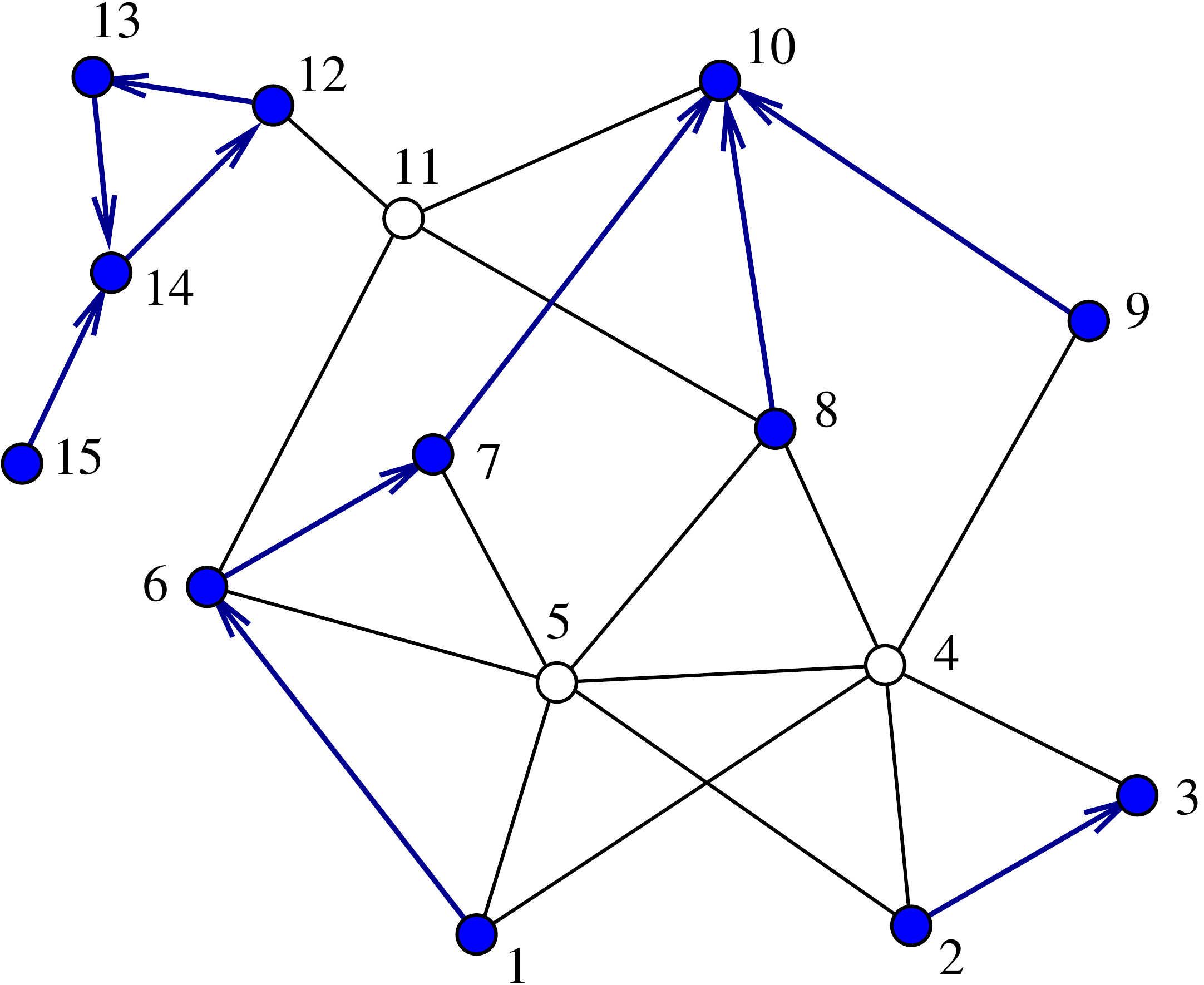}
\end{center}
\caption{\label{fig:config}
Graphical representation of a
 microscopic configuration $\{A_1=6, A_2=3, A_3=3, A_4=0,
A_5=0, A_6=7, A_7=10, A_8=10, A_9=10, A_{10}=10, A_{11}=0,
A_{12}=13, A_{13}=14, A_{14}=12, A_{15}=14\}$
for a small graph with $N=15$ vertices. A vertex $i$ is drawn as an
open circle if its state $A_i=0$,
otherwise it is drawn as a filled circle. If
the state of vertex $i$ is $A_i=j \neq i$, then we add an arrow on the
edge $(i, j)$ with this arrow pointing from $i$ to $j$.
Vertices $3$ and $10$ are two root
vertices, therefore they do not have out-going arrows.
}
\end{figure}

A microscopic configuration
of the whole graph is denoted as $\underline{A} \equiv \{A_1, A_2,
\ldots, A_N\}$, it can be represented
graphically in the following way:
If the state of a vertex $i$ is $A_i=0$, then we
represent vertex $i$ as an open circle (indicating the vertex is
un-occupied); if $A_i\neq 0$ then we represent $i$
as a filled circle (indicating the vertex is occupied); if $A_i=j\neq i$, then
we add an arrow pointing from $i$ to $j$ on the edge $(i, j)$ to indicate that
$j$ is a parent vertex of $i$. (In the case of
$A_i=i$, since $i$ is a root vertex, we do not
add any out-going arrows on the attached edges of
$i$.) Figure~\ref{fig:config} shows a
simple example of
this graphical representation.

Given a microscopic configuration $\underline{A}$,
the total number of occupied vertex, $n(\underline{A})$, and
the total number of occupied edges, $m(\underline{A})$, are
computed respectively through
\begin{eqnarray}
n(\underline{A}) & = & \sum\limits_{i=1}^N
\bigl( 1 - \delta_{A_i}^0 \bigr)\; , \\
m(\underline{A}) & = & \sum\limits_{(i, j)\in G} \bigl(1- \delta_{A_i}^0
\bigr) \bigl(1-\delta_{A_j}^0 \bigr) \; .
\end{eqnarray}
In these two expressions, $\delta_n^l$ is the
Kronecker symbol such that $\delta_n^l=1$ if $l=n$ and
$\delta_n^l=0$ if $l\neq n$.

Let us define
an edge factor $C_{i j}(A_i, A_j)$ for any edge $(i, j)$ as
\begin{eqnarray}
& & C_{i j}(A_i, A_j)  \equiv
\delta_{A_i}^0 \delta_{A_j}^0 \nonumber \\
& & \quad\quad
+\delta_{A_i}^{0} \bigl(1-\delta_{A_j}^0 - \delta_{A_j}^{i} \bigr)
+\delta_{A_j}^0 \bigl(1 - \delta_{A_i}^0 - \delta_{A_i}^j \bigr) \nonumber \\
& & \quad\quad
+ \delta_{A_i}^{j} \bigl(1- \delta_{A_j}^0 - \delta_{A_j}^{i} \bigr)
+ \delta_{A_j}^{i} \bigl(1- \delta_{A_i}^0 - \delta_{A_i}^{j} \bigr)
 \; .
 \label{eq:Cij}
\end{eqnarray}
The value of the edge factor $C_{i j}(A_i, A_j)$ is either $0$ or $1$. It is
simple to check that $C_{i j}(A_i, A_j) = 1$ in the following
five situations:
(i) $A_i = A_j = 0$ (both vertex $i$ and vertex $j$ are un-occupied);
(ii) $A_i = 0$ and $0 < A_j \neq i$ (vertex $i$ is un-occupied while vertex $j$
is occupied, and  $i$ is not the parent vertex of $j$);
(iii) $A_j=0$ and $0 < A_i \neq j$ (vertex $j$ is un-occupied while vertex $i$
is occupied, and  $j$ is not the parent vertex of $i$);
(iv) $A_i = j$ and $ 0 < A_j \neq i$ (both vertex $i$ and vertex $j$ are
occupied, and $j$ is the parent of $i$ but $i$ is not the parent of $j$);
(v) $A_j=i$ and $0 < A_i \neq j$ (both vertex $i$ and vertex $j$ are
occupied, and $i$ is the parent of $j$ but $j$ is not the parent of $i$).
For all the other input values of $A_i$ and $A_j$
the value of $C_{i j}(A_i, A_j)$ is zero.

In this work we regard each edge $(i, j)$ of the graph $G$ as a
local constraint to the microscopic configurations. Given a microscopic
configuration $\underline{A}$,
an edge $(i, j)$ is regarded as being satisfied if
 $C_{i j}(A_i, A_j) = 1$, otherwise it is regarded as being unsatisfied.
If a microscopic configuration $\underline{A}$ satisfies all the edges of
the graph $G$, it is then referred to as a  \emph{solution} of
this graph.
Figure ~\ref{fig:config} shows a solution
$\underline{A}$ for a small graph of $N=15$ vertices. Under our
graphical representation, the occupied vertices of this solution
form three connected components, the component formed by the
set of vertices $\{2, 3\}$ and the other component formed by the
set of vertices  $\{1,6,7,8,9,10\}$
are both free of any cycles (they are trees), while the
component formed by the set of vertices $\{12, 13,14,15\}$ contains
a single cycle.

A tree subgraph has $n\geq 1$
vertices and $n-1$ edges.
In the following discussions, we refer to a connected
subgraph with a single cycle as a \emph{c-tree}. By definition
a c-tree has $n\geq 3$ vertices and $n$ edges.
It can be easily proven that, in general, the occupied vertices of
any solution $\underline{A}$ of a graph $G$ form a subgraph with one
or more connected components, with each connected component being
either a tree or a c-tree. In the following discussions we refer to
such subgraphs of $G$ as the legitimate subgraphs and generically
denote them as $G_T$.

The solutions of the graph $G$ are closely related to the feedback
vertex sets of this graph. Suppose $\underline{A}$ is a solution of
$G$, then the occupied vertices of this solution form a subgraph of
disjoint trees and c-trees. Each c-tree has exactly one cycle, and the
cycles of different c-trees are mutually
disconnected. We can randomly delete one vertex from each
of these single cycles to turn a c-tree into a tree
(or a forest if the deleted vertex has more than two neighbors in the
c-tree). After this
deletion process the resulting subgraph must be free
of any cycles, therefore all the vertices not belonging to this
subgraph form a FVS.
Notice that if the occupied vertices of the solution $\underline{A}$
form an extensive number of c-trees, the size of the FVS obtained from
$\underline{A}$ will be extensively larger than the number of
un-occupied vertices in $\underline{A}$.

On the other hand, for each cycle-free subgraph of
graph $G$, we can randomly assign one vertex (say $i$) of each
connected tree component of this subgraph as the root vertex (i.e.,
setting $A_i=i$), then there is a unique way of fixing the state
variable $A_j$ of all
the other vertices $j$ of this tree component.
By repeating this assigning process for all the tree components of
this cycle-free subgraph and then fixing all the other vertices $k$ not
belonging to  this subgraph to the un-occupied state $A_k=0$,
we obtain a solution $\underline{A}$ for the graph $G$.

Consider a legitimate
subgraph $G_{T}$ of the graph $G$ which is formed by trees and
c-trees. Is there a one-to-one correspondence between $G_T$ and
a solution $\underline{A}$ of the graph $G$? The answer is no. Each
subgraph $G_T$ corresponds to many solutions of graph $G$. To explain
this, let us assume $G_T$ is composed of a non-empty set of trees
and a non-empty set of c-trees. For each tree $t \in
G_T$, we can randomly choose a vertex of this tree as the root vertex
and then fix all the other vertices. The total number of
different configurations for this tree is therefore equal to the
total number $|t|$ of vertices in tree $t$. For each c-tree
$c \in G_T$ there are two ways of fixing the arrow directions for the
edges on the cycle, therefore the total number of
different configurations for this c-tree is simply $2$. From
these discussions we know that each legitimate subgraph $G_T$
of the graph $G$ corresponds to
\begin{equation}
\mathcal{C}(G_T) \equiv 2^{n_c(G_T)} \prod\limits_{{\rm tree}\ t\in G_T} |t|
\end{equation}
different solutions $\underline{A}$ of $G$, where $n_c(G_T)$ is the
total number of c-trees in the subgraph $G_T$. The number
$\mathcal{C}(G_T)$ can be regarded as the degree of degeneracy of
the legitimate subgraph $G_T$.

After we have defined a state variable for each vertex,
we can define a partition function for the system as
\begin{equation}
\label{eq:Zx}
Z(x) = \sum\limits_{\underline{A}}
\exp\Bigl[ x \sum\limits_{i=1}^{N} (1- \delta_{A_i}^{0} ) w_i
\Bigr]
\prod\limits_{(i,j)\in G} C_{i j}(A_i, A_j) \; ,
\end{equation}
where $w_i \geq 0$ is the fixed weight of each vertex $i$,
and $x$ is a positive re-weighting parameter.
Due to the product term of
edge factors, only microscopic
configurations satisfying all the edges of $G$ have non-zero
contributions to the partition function. The re-weighting
parameter $x$ favors microscopic configurations with more
occupied vertices and larger total weights.

The partition function can also be expressed as a sum over all the
legitimate subgraphs $G_T$:
\begin{equation}
Z(x) = \sum\limits_{G_T}
\mathcal{C}(G_T) \exp\bigl[ x W( G_T ) \bigr] \; ,
\end{equation}
where $W( G_T ) \equiv \sum_{i\in G_T} w_i$ means the total weight
of vertices in the subgraph $G_T$.
Notice that, for two legitimate subgraphs
$G_T$ and $G_T^\prime$ of identical total weight $W$, their
contributions to the partition function will be different if
$\mathcal{C}(G_T) \neq \mathcal{C}(G_T^\prime)$.
In other words, the partition function $Z(x)$ does not
weight uniformly all the legitimate subgraphs of the same total weight
$W$ but favors those legitimate subgraphs $G_T$ with larger degrees of
degeneracy $\mathcal{C}(G_T)$.
 We are not much worried by this bias
issue, since the minimum FVS problem corresponds to the
$x\gg 1$ limit of our partition function. At the
limit of large $x$, the partition function
$Z(x)$ is contributed exclusively by the legitimate
subgraphs of maximum total weight, and the small differences
among the degrees of degeneracy of these subgraphs
become unimportant.

Let us define the free entropy $\Phi(x)$ of the spin glass system as
\begin{equation}
\Phi(x) = \frac{1}{x} \ln Z(x) \; .
\end{equation}
For a graph $G$ containing a large number $N$ of vertices, we expect
the free entropy $\Phi(x)$ to be an extensive thermodynamic quantity,
namely $\Phi(x) \simeq N \phi(x)$. The free entropy density
$\phi(x)$ does not depend on $N$ in the thermodynamic limit of
$N\rightarrow \infty$.

\section{Replica-symmetric mean field theory}
\label{sec:RS}

\begin{figure}
  \begin{center}
    \includegraphics[width=0.45\textwidth]{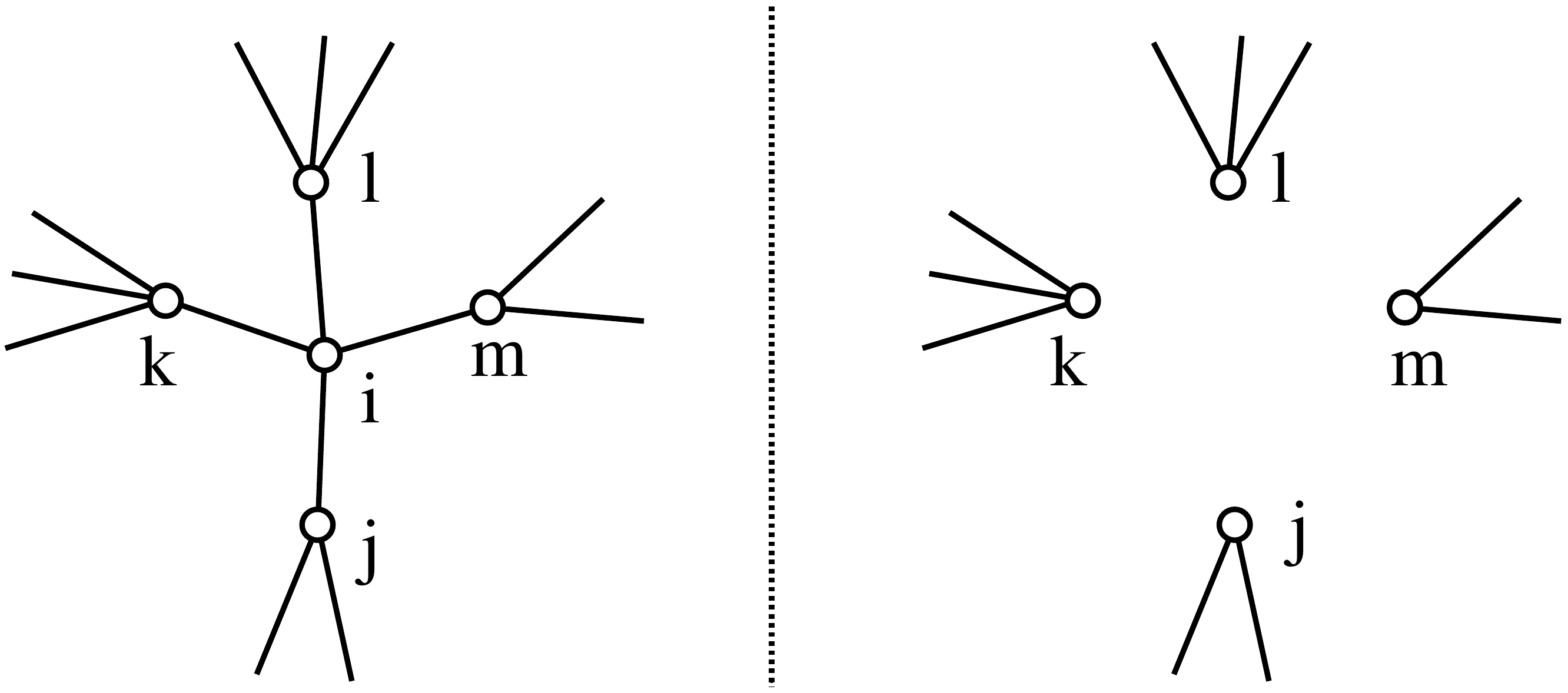}
  \end{center}
  \caption{
    \label{fig:BPapprox}
    A simple explanation on the Bethe-Peierls approximation. The
    central vertex $i$ on the left panel is connected to several
    other vertices ($\partial i = \{j, k, m, l\}$ in this
    example). Vertex $i$ mediates strong
    correlations among the states of these neighboring vertices.
    If vertex $i$ is removed from the
    graph (right panel), these neighboring vertices no longer
    feel the common effect from vertex $i$ but they may still be
    correlated due to other remaining paths of the graph. As a
    simplest approximation we ignore all the possible remaining
    correlations and assume that the vertices in set
    $\partial i$ are mutually independent of each other
    when vertex $i$ is removed.
  }
\end{figure}

Consider a randomly chosen vertex $i$ of the graph $G$,
and denote by $q_i^{A_i}$ the marginal probability that this
vertex takes the state $A_i$. The vertex $i$ may be
connected to some other vertices of the graph (see
for example the left panel of Figure~\ref{fig:BPapprox}), and its state
$A_i$ is then influenced greatly by the states of these
neighboring vertices. In return the states of the vertices in
the neighboring vertex set $\partial i$ are also strongly
influenced by the state $A_i$ of vertex $i$. To avoid over-counting
in computing the marginal probability $q_i^{A_i}$ of vertex $i$,
it is helpful for us to first remove vertex $i$ from the graph
and consider all the possible vertex state combinations of the
set $\partial i$ in the remaining system (referred to as a
cavity graph, see the right panel of Figure~\ref{fig:BPapprox}). In this
cavity graph the vertices of set $\partial i$ might still be
correlated, but in our mean field treatment we neglect all these
possible correlations and assume independence of probabilities.
This approximation is commonly known as the Bethe-Peierls
approximation
\cite{Bethe-1935,Peierls-1936,Peierls-1936a,Mezard-Montanari-2009} in
the statistical physics community.

Let us denote by
$P_{\backslash i}(\{A_j : j\in \partial i\})$ as the state joint probability
distribution of  the neighboring vertices of vertex $i$ in the cavity
graph (where vertex $i$ has been removed). In our mean field treatment
this joint probability distribution is then approximated by the
following
factorized form:
\begin{equation}
  \label{eq:BPapprox}
  P_{\backslash i}(\{A_j : j\in \partial i\}) \approx
  \prod\limits_{j\in \partial i} q_{j\rightarrow i}^{A_j} \; ,
\end{equation}
where $q_{j\rightarrow i}^{A_j}$ denotes the marginal probability
distribution
of the state $A_j$ of vertex $j \in \partial i$ in the cavity graph,
where the effect of vertex $i$ is not considered.

If all the vertices $j \in \partial i$  are either empty ($A_j=0$) or
are roots ($A_j=j$) in the cavity graph, then vertex $i$ can be a root
($A_i=i$) when it is added to the graph. This is because a
neighboring vertex $j$ can adjust its state to $A_j = i$ after
vertex $i$ is added even if
its state is $A_j=j$ in the cavity graph.
Similarly, if one vertex $l \in \partial
i$ is occupied in the cavity graph and all the other vertices of set
$\partial i$ are either empty or are roots in the cavity graph, then
vertex $i$ can take the state $A_i=l$ when it is added to the graph.
These considerations, together with the Bethe-Peierls approximation
(\ref{eq:BPapprox}), lead to the following expressions for the
marginal probability $q_i^{A_i}$:
\begin{eqnarray}
q_i^0 & = &
\frac{1}{z_i} \; , \\
q_i^i & = &
\frac{e^{x w_i} \prod\limits_{j\in \partial i}\bigl( q_{j\rightarrow i}^{0}
+ q_{j\rightarrow i}^{j} \bigr)}{z_i }
 \; , \\
q_i^l & = &
\frac{e^{x w_i} (1-q_{l\rightarrow i}^{0})
\prod\limits_{k\in \partial i\backslash l}\bigl(q_{k\rightarrow i}^0+
q_{k\rightarrow i}^{k} \bigr)}{z_i}
\; , \quad l\in \partial i \; \nonumber \\
& &
\end{eqnarray}
where the normalization constant $z_i$ is calculated by
\begin{eqnarray}
z_i  &\equiv & 1 + e^{x w_i} \times
\biggl[\prod\limits_{j\in \partial i}\bigl( q_{j\rightarrow i}^{0}
+ q_{j\rightarrow i}^{j} \bigr) + \nonumber \\
& & \quad
\sum\limits_{j\in \partial i} (1-q_{j\rightarrow i}^{0})
\prod\limits_{k\in \partial i\backslash j}\bigl(q_{k\rightarrow i}^0+
q_{k\rightarrow i}^{k} \bigr) \biggr] \; .
\end{eqnarray}
In the above expressions, $\partial i\backslash j$ means the set of
all the neighboring vertices of vertex $i$ except vertex $j$.

After the marginal probabilities $q_i^{A_i}$ for all the vertices $i$
have been obtained,
the mean fraction of occupied vertices $\rho$ is easily calculated through
\begin{equation}
  \rho = 1 - \frac{1}{N} \sum\limits_{i=1}^{N} q_i^0 \; ,
\end{equation}
and the relative total weight of the occupied vertices $\omega$ is
obtained through
\begin{equation}
  \omega \equiv \frac{1}{N} \sum\limits_{i=1}^{N} (1-q_i^0) w_i \; .
\end{equation}

Under the Bethe-Peierls approximation
the free entropy $\Phi(x)$ has the following simple expression:
\begin{equation}
  \label{eq:Phi}
  \Phi(x) = \sum\limits_{i=1}^{N} \phi_i - \sum\limits_{(i,j)\in G}
  \phi_{i j} \; ,
\end{equation}
where $\phi_i$ and $\phi_{i j}$ are, respectively, the free entropy
contribution of a vertex $i$ and an edge $(i,j)$:
\begin{eqnarray}
 \hspace*{-0.5cm}\phi_i & = & \frac{1}{x}
  \ln \biggl[ 1 + e^{x w_i} \prod\limits_{j\in \partial i}
    [ q_{j\rightarrow i}^0 + q_{j\rightarrow i}^j ] + \nonumber \\
    & & \quad
    e^{x w_i} \sum\limits_{j\in \partial i} (1- q_{j\rightarrow i}^0 )
    \prod\limits_{k\in \partial i\backslash j} (q_{k\rightarrow i}^0 +
    q_{k\rightarrow i}^k)
    \biggr] \; , \\
 \hspace*{-0.5cm}\phi_{i j} & = &
  \frac{1}{x} \ln \biggl[ q_{i\rightarrow j}^0 q_{j\rightarrow i}^0
    + (1-q_{i\rightarrow j}^0)(q_{j\rightarrow i}^0+q_{j\rightarrow i}^j)
    + \nonumber\\
& & \quad (1-q_{j\rightarrow i}^0)
(q_{i\rightarrow j}^0 + q_{i\rightarrow j}^i)
    \biggr] \; .
\end{eqnarray}
The free entropy expression (\ref{eq:Phi}) can be rigorously
justified from the mathematical framework of partition function
expansion \cite{Zhou-Wang-2012,Xiao-Zhou-2011,Zhou-etal-2011} or
through the cluster variation method \cite{Kikuchi-1951,An-1988}.
From (\ref{eq:Phi}) the free entropy density is then
obtained as $\phi(x) = \frac{1}{N} \Phi(x)$.
The entropy density $s$ of the system is then calculated through
\begin{equation}
  s =  x(\phi - \omega) \; .
\end{equation}

To complete the mean field theory we also need a set of equations for
the probability distributions $q_{i\rightarrow j}^{A_i}$.
Since $q_{i\rightarrow j}^{A_i}$ has the same meaning as $q_{i}^{A_i}$
but is defined on the cavity graph where vertex $j$ is being removed,
we can write down the following equations under the Bethe-Peierls
approximation:
\begin{eqnarray}
  q_{i\rightarrow j}^0 & = &
  \frac{1}{z_{i\rightarrow j} }
  \; , \\
  q_{i\rightarrow j}^i & = &
  \frac{e^{x w_i}
    \prod\limits_{k\in \partial i\backslash j}\bigl( q_{k\rightarrow i}^{0}
    + q_{k\rightarrow i}^{k} \bigr)}{z_{i\rightarrow j} }
   \; , \\
  q_{i\rightarrow j}^l & = &
  \frac{e^{x w_i} (1-q_{l\rightarrow i}^{0})
    \prod\limits_{m\in \partial i\backslash j,l}\bigl(q_{m\rightarrow i}^0+
    q_{m\rightarrow i}^{m} \bigr)}{z_{i\rightarrow j} }
  \; , \quad l\in \partial i\backslash j \; \nonumber \\
  & &
\end{eqnarray}
where  $\partial i\backslash j,l$ means the set of
all the neighboring vertices of vertex $i$ except vertex $j$ and
vertex $l$, and the normalization constant
$z_{i\rightarrow j}$ is expressed as
\begin{eqnarray}
\hspace*{-0.5cm}z_{i\rightarrow j}  &\equiv &
1 + e^{x w_i}
    \biggl[\prod\limits_{k\in \partial i\backslash j}\bigl( q_{k\rightarrow i}^{0}
      + q_{k\rightarrow i}^{k} \bigr) + \nonumber \\
      & &
       \sum\limits_{k\in \partial i\backslash j} (1-q_{k\rightarrow i}^{0})
      \prod\limits_{m\in \partial i\backslash j,k}\bigl(q_{m\rightarrow i}^0+
      q_{m\rightarrow i}^{m} \bigr) \biggr] \; .
\end{eqnarray}
These self-consistent equations are commonly
referred to as a set of belief propagation (BP) equations in the
literature.

The BP equations and the free entropy expression (\ref{eq:Phi}) form
the replica-symmetric (RS) mean field theory of the spin glass model
(\ref{eq:Zx}).
For a single graph instance $G$, we can iterate the BP equations
on the edges of the graph at a fixed value of re-weighting parameter
$x$.  If the BP equations are able to converge
to a fixed point, we can then calculate the entropy density $s$,
the occupation density $\rho$ and the relative total weight of
occupied vertices $\omega$ at this fixed point. The value $1-\rho$
is then the fraction of un-occupied vertices estimated by the RS
mean field theory. Because some occupied vertices of the c-trees need
to be included into the FVS besides all the un-occupied vertices, this
fraction $1-\rho$ is regarded as a lower-bound on the fraction of
vertices in the FVS.

The RS mean field theory can also be used to calculate
ensemble-averaged properties. Let us first consider the ensemble of
finite-connectivity Erd\"os-R\'enyi (ER) random graphs.
Such an ensemble is characterized by a mean vertex degree $c$ and
a Poisson degree distribution
\begin{equation}
  \label{eq:Poisson}
  P(d) = \frac{e^{-c} c^d}{d!} \; ,
\end{equation}
which gives the probability that a randomly chosen vertex $i$ has $d$
edges attached \cite{He-Liu-Wang-2009}.
We create a large population array of two-dimensional elements
$(q_{i\rightarrow j}^0, q_{i\rightarrow j}^i)$
to represent the messages on all the edges of
a random graph. This population array is then updated until
the distribution of elements in the array no longer changes
with time. We then keep updating the population to
compute through the mean field expressions
the thermodynamic quantities such as $\rho$, $\omega$, $\phi$,
and $s$. For simplicity we set the weight $w_i$ of each
vertex $i$ to be $w_i=1$ in all our following numerical calculations.

In each step of the above-mentioned population updating process,
first an integer value $d$ is generated according to the Poisson
distribution (\ref{eq:Poisson}). This value $d$ is considered as the
degree of a central vertex, say $i$. We then randomly choose $d$ elements
from the population array and consider them as the input messages
$(q_{j\rightarrow i}^0, q_{j\rightarrow i}^j)$ from the $d$
neighboring vertices $j$ of vertex $i$. Then we obtain $d$ new output
messages $(q_{i\rightarrow j}^0, q_{i\rightarrow j}^i)$ according to
the BP equations and replace $d$ randomly chosen elements of the
population array by these $d$ new ones. Such a kind of population dynamics
simulations is now commonly used  for studying the
ensemble-averaged properties of spin glasses, see, for example, the
textbook \cite{Mezard-Montanari-2009}.

\begin{figure}
  \begin{center}
    \subfigure[]{
      \label{fig:EntropyRSpop10:a}
      \includegraphics[width=0.45\textwidth]{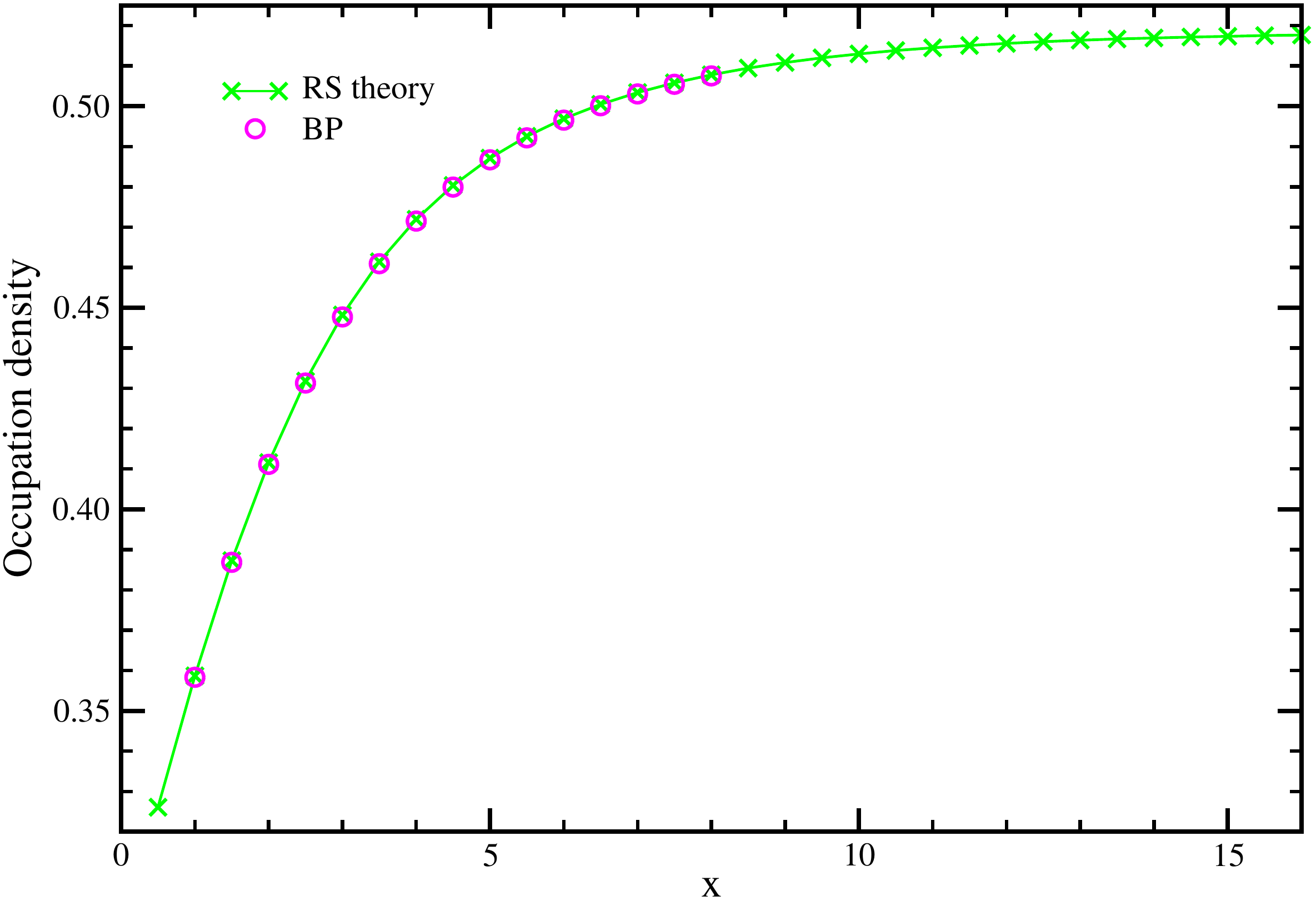}
    }
    \vskip 0.1cm
    \subfigure[]{
      \label{fig:EntropyRSpop10:b}
      \includegraphics[width=0.45\textwidth]{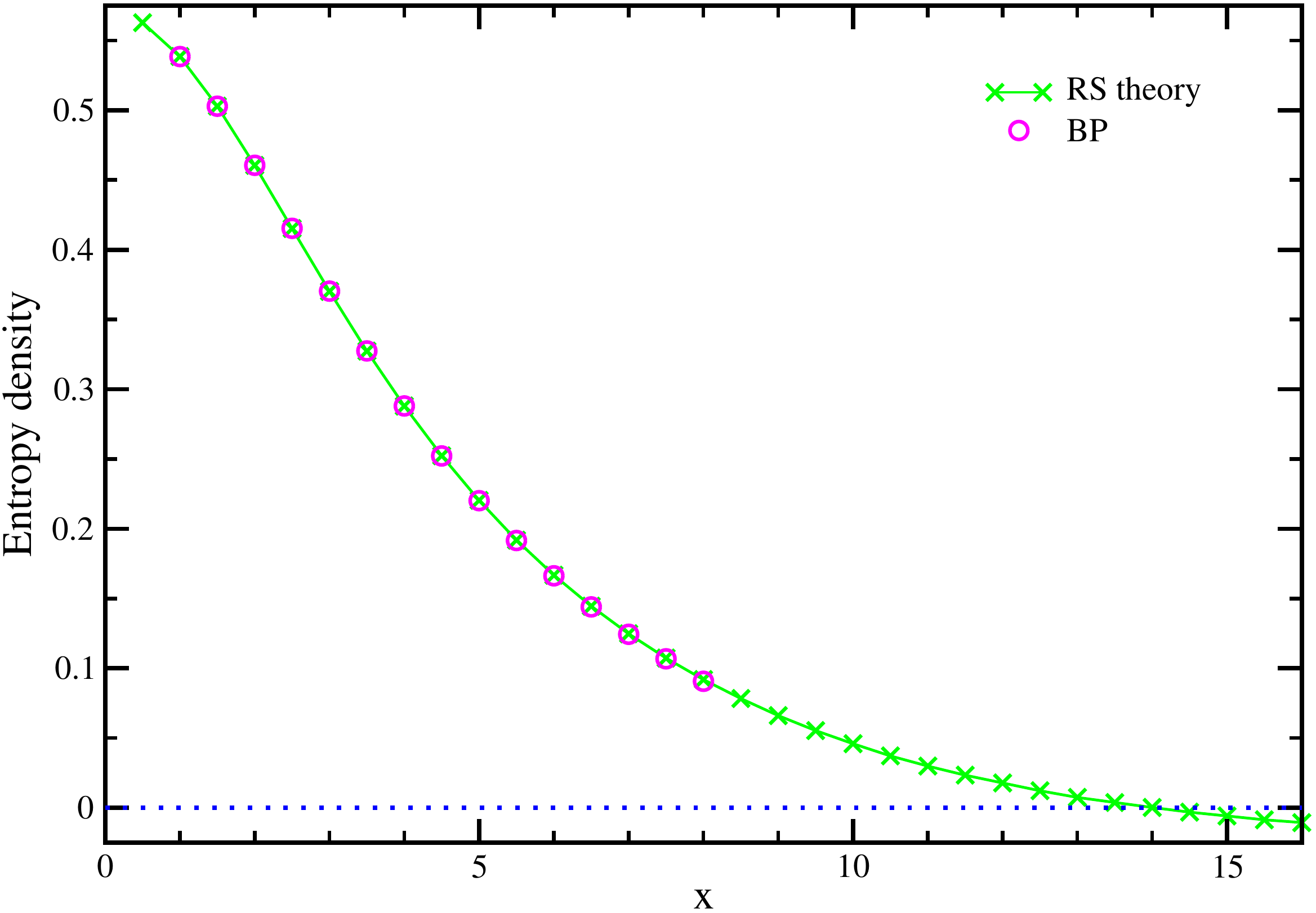}
    }
    \vskip 0.1cm
    \subfigure[]{
      \label{fig:EntropyRSpop10:c}
      \includegraphics[width=0.45\textwidth]{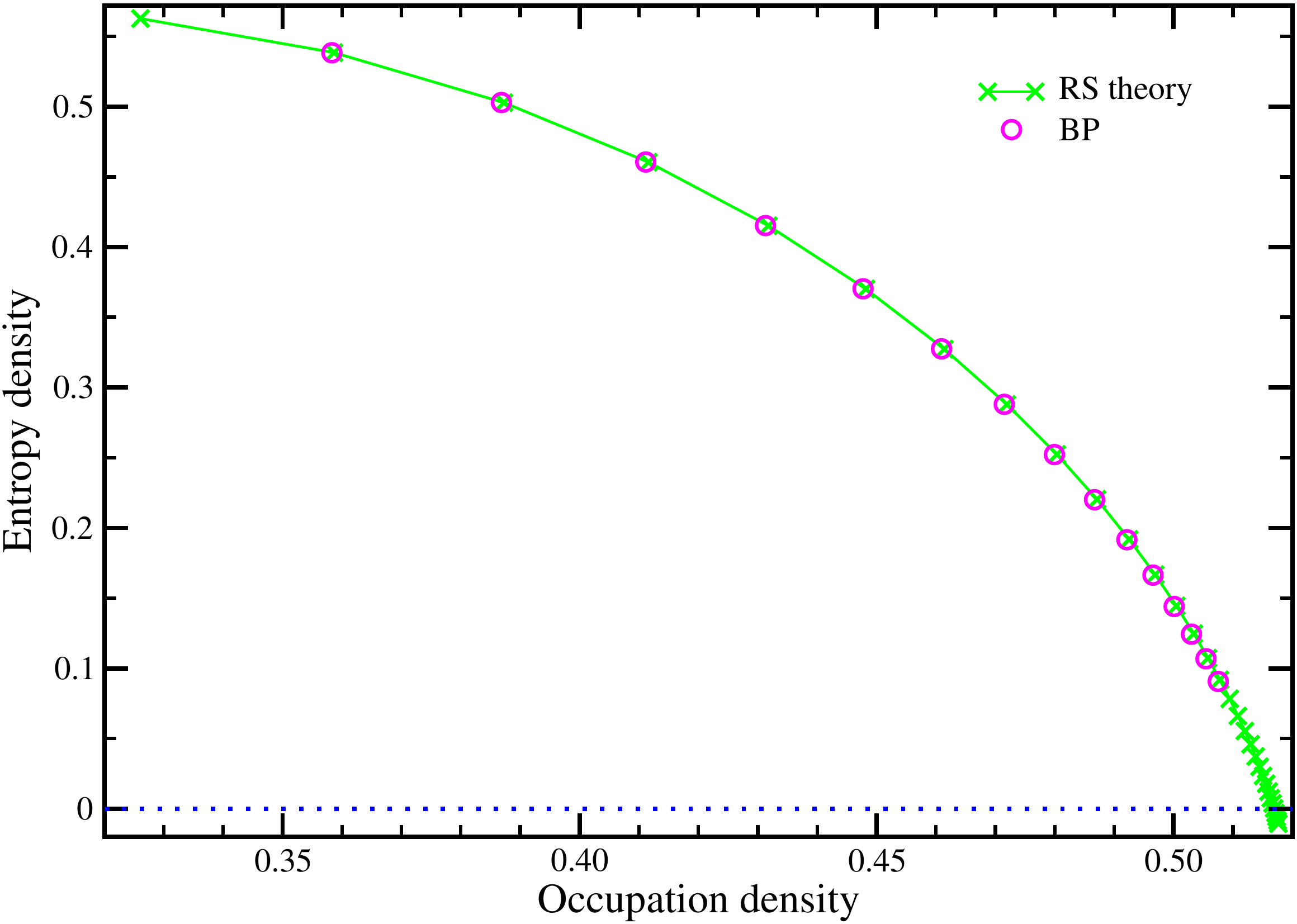}
    }
  \end{center}
    \caption{\label{fig:EntropyRSpop10}
      Replica-symmetric mean field results on
      ER random graphs of mean degree $c=10$. Cross symbols are
      ensemble-averaged results, while circles are results obtained by
      BP iteration on a single random graph instance of $N=10^5$ vertices.
      (a) Mean occupation density $\rho$. (b) Entropy density $s$.
      (c) Entropy density $s$ as a function of occupation density
      $\rho$ obtained by eliminating the re-weighting parameter
      $x$ from each pair of points $(x, \rho)$ and $(x, s)$
      of (a) and (b). The dotted lines of (b) and (c) indicate $s=0$.
      BP iteration fails to converge at $x\geq 8$ on the single
      graph instance.
    }
\end{figure}

Figure \ref{fig:EntropyRSpop10} shows the mean field
results for the ER random graph ensemble with mean degree
$c=10$. The occupation density $\rho$ increases
with re-weighting parameter $x$ (Figure~\ref{fig:EntropyRSpop10:a}),
while the entropy density $s$
decreases with $x$ and becomes negative at $x > 14$
(Figure~\ref{fig:EntropyRSpop10:b}).
The entropy density $s$ as a function of occupation
density $\rho$ is shown in Figure~\ref{fig:EntropyRSpop10:c}, which
appears to be concave.

If the entropy density $s$ is positive even at $x \rightarrow \infty$,
we take the value of $\rho = \rho_0$ at $x\rightarrow \infty$ as the
maximal
occupation density the system can achieve. On the other hand, if
the calculated entropy density $s$ becomes negative at large values of $x$,
since the true entropy density of a spin glass system with discrete state
variables should be non-negative, the point $\rho=\rho_0$
at which $s(\rho_0)=0$ is regarded as the maximum value of occupation
density the system can achieve.

In random graphs, since the typical cycle length diverges logarithmically with the vertex number $N$, the correction effect of the single cycle of
each c-tree to the FVS size will be of order at most $[\ln N]^{-1}$.
Therefore these correction effects can be safely neglected in the
thermodynamic limit of $N\rightarrow \infty$.
The fraction of vertices in the
minimum feedback vertex sets is then obtained as $1-\rho_0$ for the
random graph ensemble.

At mean degree $c=10$ the mean field results of
Figure \ref{fig:EntropyRSpop10} suggest that $\rho_0
\approx 0.517$, namely each minimum FVS contains about
$0.483 N$ vertices of the random graph. The minimum FVS size as predicted
by the RS mean field theory is shown in Figure~\ref{fig:ERfvs} as a
function of mean vertex degree $c$ (the
cross symbols). As expected, the minimum FVS size
increases continuously with $c$.

We can also perform BP simulations on single
random graph instances. A single ER random graph instance can be
easily generated by the following way:
start from an empty graph of $N$ vertices and
zero edges, we keep adding an edge to two randomly
chosen different vertices until the
total number of edges in the graph reaches
$M= (c/2) N$ (of cause, self-connections
and multiple edges between the same pair of vertices are discarded).
For such a large single random graph instance, we find that if the
BP iteration process is able to converge to a fixed point,
the occupation density $\rho$ and the
entropy density $s$ calculated at this fixed point coincide
with the ensemble-averaged values.
However, if the mean degree $c >= 4$ and the re-weighting parameter $x$ is
large, the BP iteration process fails to converge to a fixed point.
For example, in the case of $c= 10$ our preliminary results
suggest that BP iteration is not convergent when $x\geq 8$ (see
Figure~\ref{fig:EntropyRSpop10}).

The non-convergence of BP on single random graph instances (with
mean degree $c\geq 4$) at large values of $x$
indicates that the RS mean field theory is not sufficient
to describe the FVS problem at high occupation densities.
We need to consider correlations among the states of the neighboring
vertices of each given vertex $i$, and the Bethe-Peierls approximation
Eq.~(\ref{eq:BPapprox}) has to be improved. This can be achieved by
applying the first-step replica-symmetry-breaking (1RSB) mean field theory
\cite{Mezard-Parisi-2001,Zhou-Wang-2012,Xiao-Zhou-2011,Zhou-etal-2011}.
We will return to this issue and the related spin glass phase transition
problem in a future paper.

We also work on the ensemble of regular random  graphs. In a regular
random  (RR) graph, each of the vertices has exactly $K$ edges but the
graph is otherwise completely random. The RS mean field predictions on
the minimum FVS size of this RR graph ensemble are shown in
Figure~\ref{fig:RRfvs}. At each value of degree $K$ the RS prediction
slightly exceeds the mathematical lower-bound obtained by
Bau and co-authors \cite{Bau-Wormald-Zhou-2002}.

\section{Belief propagation-guided decimation}
\label{sec:BPD}

The RS mean field theory can also guide us to construct feedback
vertex sets for single graph instances. We have implemented
a simple belief propagation-guided
decimation (BPD) algorithm as follows:
\begin{enumerate}
\item[(0).]
  Input a graph G and initialize randomly the edge messages
  $(q_{i\rightarrow j}^0, q_{i\rightarrow j}^i)$ and
  $(q_{j\rightarrow i}^0, q_{j\rightarrow i}^j)$ for each
  edge $(i, j)$ of the graph $G$.
  The feedback vertex set $\Gamma$ is
  initialized to be empty. The re-weighting parameter
  $x$ is set to an appropriate value (e.g., $x\approx 10$).

\item[(1).]
  Perform the BP iteration process a number $T$ of rounds
  (in each round of the  iteration, the vertices of the graph
  $G$ are randomly ordered and their output messages are
  then updated sequentially).  A fixed point of BP equations may not be
  reached after these $T$ rounds of iteration. No matter whether a BP
  fixed point has reached, we compute the empty probability $q_i^0$ of
  each vertex $i$ based on the current inputting
  messages to vertex $i$. Then the $f N$ vertices with the highest
  empty probability values are added to the set $\Gamma$,  and these
  vertices are then
  removed from the graph $G$ together with all the edges attached to them.

\item[(2).]
  Then we further simplify the graph $G$ by recursively removing vertices of degree $0$ or $1$ until all the remaining vertices of the graph
  have two or more attached edges. Notice that these removed
  vertices are \emph{not}   added to
  the set $\Gamma$.

\item[(3).]
  If the graph $G$ is non-empty, we repeat the above-mentioned
  step (1) and step (2).

\item[(4).]
  Output the resulting set $\Gamma$.
\end{enumerate}
During the decimation process, if the remaining graph still contains cycles, at least
one vertex will be moved to the set $\Gamma$ to decrease the number of cycles.
The BPD process will terminate only when no cycles are present in the remaining graph.
Therefore the set $\Gamma$ is a feedback vertex set
of $G$. In
other words, the subgraph of $G$ obtained by removing all the vertices of
$\Gamma$ is a forest (there are usually many
tree components in this forest but no c-trees).

\begin{figure}
\begin{center}
    \subfigure[]{
      \label{fig:ERfvs}
      \includegraphics[width=0.45\textwidth]{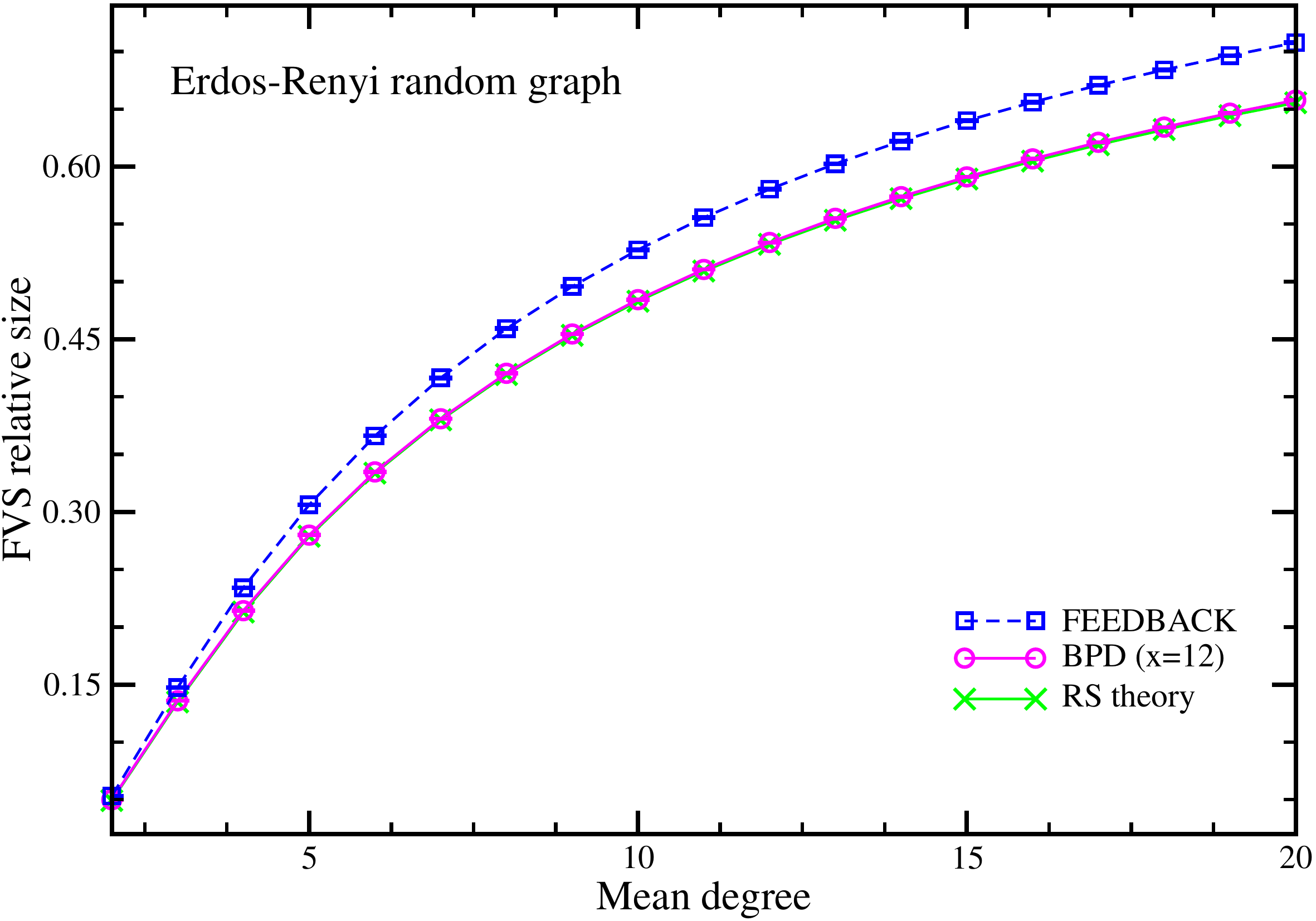}
    }
    \vskip 0.1cm
    \subfigure[]{
      \label{fig:RRfvs}
      \includegraphics[width=0.45\textwidth]{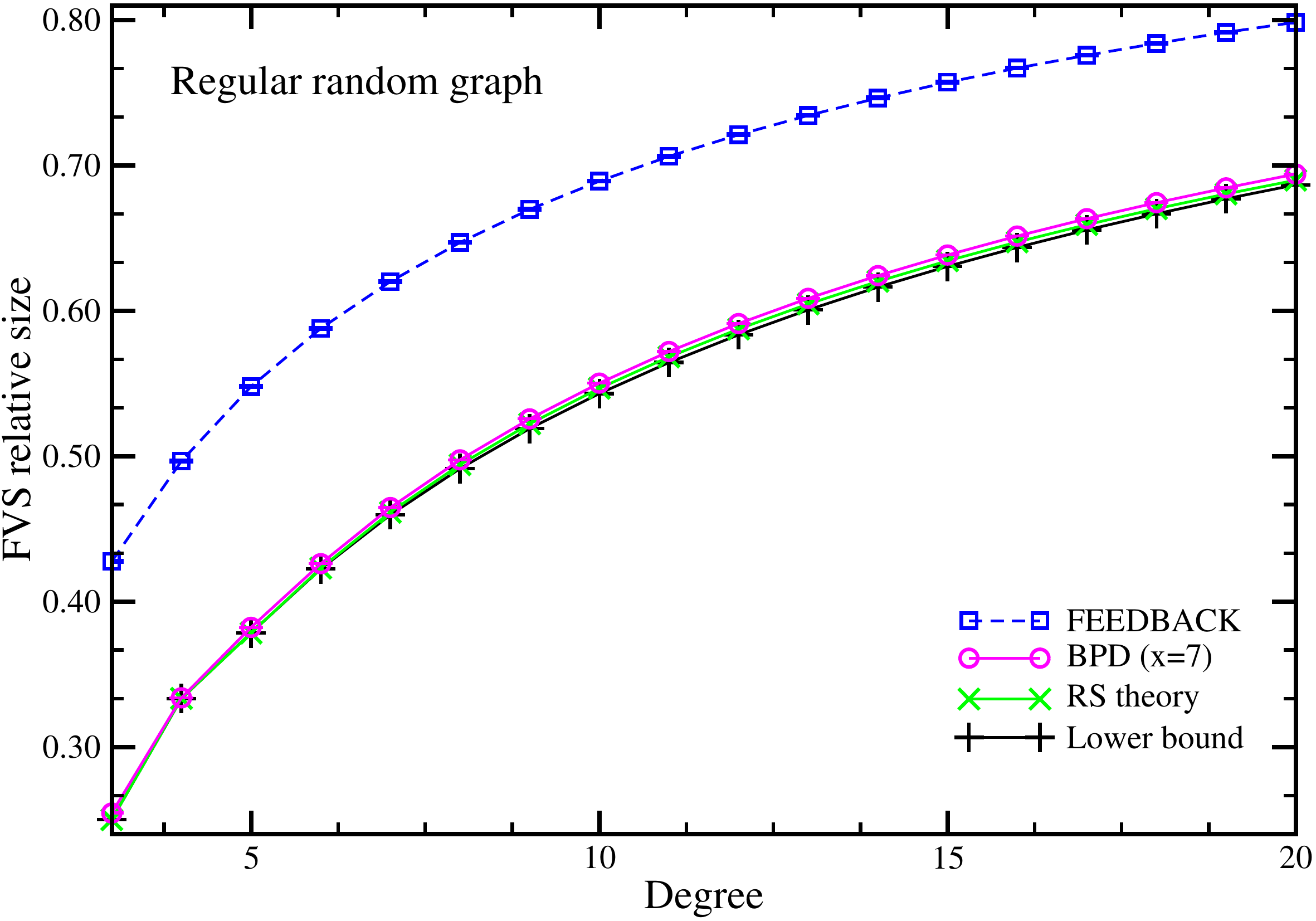}
    }
  \end{center}
  \caption{\label{fig:ER_RR_fvs}
    Comparing the theoretical predictions and algorithmic results
    on the minimum FVS sizes. (a) Erd\"os-R\'enyi random graphs;
    (b) regular random graphs. Cross symbols are the RS mean field
    predictions; circle symbols (together with error bars) are
    the average values of the FVS sizes obtained by a
    single run of the BPD algorithm on $96$ random graph instances
    of $N=10^5$ vertices; square symbols (together with error bars)
    are the average values of the FVS sizes obtained by
    a single run of the FEEDBACK algorithm
    \cite{Bafna-Berman-Fujito-1999} on the same $96$ random
    graph instances. The re-weighting parameter of the BPD is fixed
    to $x=12$ in the case of ER graphs, and to $x=7$ in the
    case of RR graphs.
    The mathematical lower-bounds on the FVS size of RR graphs (see
    the third column of Table 2 in \cite{Bau-Wormald-Zhou-2002})
    are shown as plus symbols in (b).
  }
\end{figure}

We have implemented the above BPD algorithm using {\tt C++}
programming language (the code is freely available upon request).
In our numerical simulations we set the
BPD parameters to be $T=500$ and $f=0.01$. These parameters
are not necessarily optimal but are chosen so that a single run
of the BPD algorithm on a large graph instance of $N=10^5$ vertices
and $M=10^6$ edges will terminate within three to four hours.
If the fraction $f$ is further reduced, say to $f=0.001$, then
the BPD algorithm will reach slightly smaller feedback vertex sets, but
the computing time is much longer.

We have tested the performance of the BPD algorithm at
different fixed values of the re-weighting parameter $x$.
The sizes of the constructed feedback vertex sets $\Gamma$
only change very slightly with different choices of $x$. For ER random
graphs the value of $x = 12$ seems to be close to optimal, while
for regular random graphs the value is $x=7$.

The results of this BPD algorithm on ER and RR graphs
are shown in Figure~\ref{fig:ERfvs} and ~\ref{fig:RRfvs},
 respectively (the circle symbols). As a comparison we
also show in the same figure the results obtained by
the well-known
FEEDBACK algorithm of Bafna and
co-workers \cite{Bafna-Berman-Fujito-1999} (the square
symbols). The FEEDBACK is a
fast heuristic algorithm that is guaranteed to construct a FVS
of size not exceeding two times that of an optimal FVS.

We can clearly see from Figure~\ref{fig:ER_RR_fvs} that the
sizes of feedback vertex sets constructed by
the BPD algorithm reach the predicted minimum FVS sizes of
the RS mean field theory. On the other hand, for a given random graph
instance, the feedback vertex sets
constructed by the FEEDBACK algorithm are extensively larger in size
than those
constructed by the BPD algorithm. The good agreement between the results of
the BPD algorithm and the mean field predictions indicates that the BPD
algorithm is excellent for random graph instances, and it also indicates
that the RS mean field theory is very good in predicting the mean
minimum FVS sizes of random graphs (the predictions can be further
improved slightly if ergodicity-breaking is considered in the theory).

We have also applied the BPD algorithm on
hyper-cubic regular lattices with
periodic boundary conditions.
For two-dimensional
square lattices the feedback vertex sets
obtained by the BPD algorithm ($x=7$)
contain about $35.1\%$ of the vertices. This value is
very close to the mathematical lower-bound of
$\frac{1}{3}$ obtained by Beineke and Vandell
\cite{Beineke-Vandell-1997,Bau-Wormald-Zhou-2002} and
is much
better than the value of $49.5\%$ obtained
by the FEEDBACK algorithm.
For three-dimensional
cubic lattices the feedback vertex sets
obtained by the BPD algorithm ($x=7$)
contain about $41.9\%$ of the vertices, which is
again very close to the mathematical lower-bound of
$\frac{2}{5}$ \cite{Beineke-Vandell-1997,Bau-Wormald-Zhou-2002}
and much better than the value of $49.9\%$ obtained
by the FEEDBACK algorithm. The performance of the
BPD algorithm may be further improved if we consider
explicitly the correlation effect of short loops in the
iteration equations (see \cite{Zhou-Wang-2012} for example).
A systematic comparison of the
performance of BPD with other optimization algorithms (such as
simulated annealing and parallel tempering) needs to be
carried out in the future.

\section{Conclusion and discussions}
\label{sec:conclusion}

We have constructed a spin glass model (\ref{eq:Zx}) for the feedback
vertex set problem on an undirected graph. We have solved this model
by replica-symmetric mean field theory on the ensemble of
finite-connectivity random graphs. We have also implemented a
belief propagation-guided decimation algorithm based on this
mean field theory and applied this algorithm to single random
graph instances and hyper-cubic regular lattices. Our numerical results
of Figure \ref{fig:ER_RR_fvs} demonstrate that the BPD message-passing
algorithm is able to construct nearly optimal feedback vertex sets for
single random graph instances and regular lattice instances.
The BPD algorithm also
has much better performance than the conventional
FEEDBACK algorithm of \cite{Bafna-Berman-Fujito-1999} when applied to
finite-dimensional hyper-cubic lattices.

Although the replica-symmetric mean field theory
appears to predict the
minimum FVS sizes of Erd\"os-R\'enyi random graphs very well, the
BP iteration process does not converge to a fixed point
on single random
graphs when the
re-weighting parameter $x$ exceeds certain threshold value.
We still need to carry out the replica-symmetry-broken mean field
calculations to fully understand the statistical physics properties
of the spin glass model (\ref{eq:Zx}) at large $x$ values.
Such a theoretical exploration is deferred to a later publication.

The FVS problem of directed graphs is even more important in
practical applications.
A way of constructing a Ising model for
the directed FVS problem has been suggested in the recent paper
of Lucas \cite{Lucas-2013}. Following the idea of
Ref.~\cite{Lucas-2013} (and also that of Ref.~\cite{Bayati-etal-2008})
we may define on each vertex $i$ of a directed graph $G$
an integer height state $h_i$ such that $h_i=0$ means vertex $i$ is
un-occupied (belonging to the FVS)
and $h_i \geq 1$ means $i$ is occupied (not belonging to the FVS).
A
height configuration of the whole system can be denoted as
$\underline{h} \equiv \{h_1, h_2, \ldots, h_N\}$.
On each directed edge $(i\rightarrow j)$
pointing from vertex $i$ to vertex $j$, a
simple edge factor $C_{i\rightarrow j}$
similar to Eq.~(\ref{eq:Cij}) can be introduced as
\begin{equation}
C_{i\rightarrow j}(h_i, h_j) = \delta_{h_j}^0 +
\bigl(1-\delta_{h_j}^0 \bigr) \Theta(h_j-h_i)
\; ,
\end{equation}
where $\Theta(n) = 0$ for integer $n\leq 0$ and
$\Theta(n) = 1$ for integer $n \geq 1$. If $h_i \times h_j=0$
then $C_{i\rightarrow j}(h_i, h_j)=1$; if $h_i \geq 1$ and
$h_j\geq 1$ (namely both $i$ and $j$ are occupied) then
$C_{i\rightarrow j}(h_i, h_j) = 1$ only if $h_i<h_j$.
A partition function
similar to Eq.~(\ref{eq:Zx}) can be defined on the directed
graph $G$ as
\begin{equation}
\label{eq:Zxd}
    Z(x) = \sum\limits_{\underline{h}}
\exp\Bigl[ x \sum\limits_{i=1}^{N} (1- \delta_{h_i}^{0} ) w_i
\Bigr]
\prod\limits_{(i\rightarrow j)\in G} C_{i\rightarrow j}(h_i, h_j) \; .
\end{equation}
Because of the product term of edge factors in
the above equation, if there is a directed cycle within the subgraph of
occupied vertices, the corresponding height configuration
$\underline{h}$ will have zero contribution to the partition function.

For the ensemble of directed ER random graphs in which each vertex on
average having $\alpha$ inputting edges and $\alpha$ out-going edges,
our preliminary RS mean field calculations indicate that at $\alpha=10.0$
a minimum feedback vertex set contains about $0.448 N$ vertices.
A detailed report of the mean field and algorithmic
 results will be presented in a later paper.

\section*{Acknowledgements}

I thank Yang-Yu Liu for introducing the feedback vertex set problem
to me and for helpful comments on the manuscript,
Victor Martin-Mayor for suggesting to work on regular lattices,
and Lenka Zdeborov\'a for suggesting to work on regular random graphs and
for pointing \cite{Bau-Wormald-Zhou-2002} to me. I also thank Ying Zeng, Chuang Wang, and Jack Raymond for helpful discussions. The main idea of this
paper emerged during a workshop organized by Lei-Han Tang
at the Beijing Computational Science Research Center  (BCSRC) in
June 2013. The hospitality of BCSRC is acknowledged.
This work was supported by the National Basic Research Program of China
(No. 2013CB932804), the Knowledge Innovation
Program of Chinese Academy of Sciences (No. KJCX2-EW-J02),
and the National Science Foundation of China (grant Nos. 11121403, 11225526).


\end{document}